\documentclass[pra,showpacs,twocolumn,superscriptaddress]{revtex4}

\def\temp{1.35}%
\let\tempp=\relax
\expandafter\ifx\csname psboxversion\endcsname\relax
\else
    \ifdim\temp cm>\psboxversion cm
    \else
      \let\temp=\psboxversion
      \let\tempp= 
    \fi
\fi
\tempp
\let\psboxversion=\temp
\catcode`\@=11
\def\psfortextures{
\def\PSspeci@l##1##2{%
\special{illustration ##1\space scaled ##2}%
}}%
\def\psfordvitops{
\def\PSspeci@l##1##2{%
\special{dvitops: import ##1\space \the\drawingwd \the\drawinght}%
}}%
\def\psfordvips{
\def\PSspeci@l##1##2{%
\d@my=0.1bp \d@mx=\drawingwd \divide\d@mx by\d@my
\includegraphics{##1\space}}}%
\def\psforoztex{
\def\PSspeci@l##1##2{%
\special{##1 \space
      ##2 1000 div dup scale
      \number-\psllx\space\space \number-\pslly\space\space translate
}}}%
\def\psfordvitps{
\def\dvitpsLiter@ldim##1{\dimen0=##1\relax
\special{dvitps: Literal "\number\dimen0\space"}}%
\def\PSspeci@l##1##2{%
\at(0bp;\drawinght){%
\special{dvitps: Include0 "psfig.psr"}
\dvitpsLiter@ldim{\drawingwd}%
\dvitpsLiter@ldim{\drawinght}%
\dvitpsLiter@ldim{\psllx bp}%
\dvitpsLiter@ldim{\pslly bp}%
\dvitpsLiter@ldim{\psurx bp}%
\dvitpsLiter@ldim{\psury bp}%
\special{dvitps: Literal "startTexFig"}%
\special{dvitps: Include1 "##1"}%
\special{dvitps: Literal "endTexFig"}%
}}}%
\def\psfordvialw{
\def\PSspeci@l##1##2{
\special{language "PostScript",
position = "bottom left",
literal "  \psllx\space \pslly\space translate
  ##2 1000 div dup scale
  -\psllx\space -\pslly\space translate",
include "##1"}
}}%
\def\psforptips{
\def\PSspeci@l##1##2{{
\d@mx=\psurx bp
\advance \d@mx by -\psllx bp
\divide \d@mx by 1000\multiply\d@mx by \xscale
\incm{\d@mx}
\let\tmpx\dimincm
\d@my=\psury bp
\advance \d@my by -\pslly bp
\divide \d@my by 1000\multiply\d@my by \xscale
\incm{\d@my}
\let\tmpy\dimincm
\d@mx=-\psllx bp
\divide \d@mx by 1000\multiply\d@mx by \xscale
\d@my=-\pslly bp
\divide \d@my by 1000\multiply\d@my by \xscale
\at(\d@mx;\d@my){\special{ps:##1 x=\tmpx cm, y=\tmpy cm}}
}}}%
\def\psonlyboxes{
\def\PSspeci@l##1##2{%
\at(0cm;0cm){\boxit{\vbox to\drawinght
  {\vss\hbox to\drawingwd{\at(0cm;0cm){\hbox{({\tt##1})}}\hss}}}}
}}%
\def\psloc@lerr#1{%
\let\savedPSspeci@l=\PSspeci@l%
\def\PSspeci@l##1##2{%
\at(0cm;0cm){\boxit{\vbox to\drawinght
  {\vss\hbox to\drawingwd{\at(0cm;0cm){\hbox{({\tt##1}) #1}}\hss}}}}
\let\PSspeci@l=\savedPSspeci@l
}}%
\newread\pst@mpin
\newdimen\drawinght\newdimen\drawingwd
\newdimen\psxoffset\newdimen\psyoffset
\newbox\drawingBox
\newcount\xscale \newcount\yscale \newdimen\pscm\pscm=1cm
\newdimen\d@mx \newdimen\d@my
\newdimen\pswdincr \newdimen\pshtincr
\let\ps@nnotation=\relax
{\catcode`\|=0 |catcode`|\=12 |catcode`|
|catcode`#=12 |catcode`*=14
|xdef|backslashother{\}*
|xdef|percentother{
|xdef|tildeother{~}*
|xdef|sharpother{#}*
}%
\def\R@moveMeaningHeader#1:->{}%
\def\uncatcode#1{%
\edef#1{\expandafter\R@moveMeaningHeader\meaning#1}}%
\def\execute#1{#1}
\def\psm@keother#1{\catcode`#112\relax}
\def\executeinspecs#1{%
\execute{\begingroup\let\do\psm@keother\dospecials\catcode`\^^M=9#1\endgroup}}%
\def\@mpty{}%
\def\matchexpin#1#2{
  \fi%
  \edef\tmpb{{#2}}%
  \expandafter\makem@tchtmp\tmpb%
  \edef\tmpa{#1}\edef\tmpb{#2}%
  \expandafter\expandafter\expandafter\m@tchtmp\expandafter\tmpa\tmpb\endm@tch%
  \if\match%
}%
\def\matchin#1#2{%
  \fi%
  \makem@tchtmp{#2}%
  \m@tchtmp#1#2\endm@tch%
  \if\match%
}%
\def\makem@tchtmp#1{\def\m@tchtmp##1#1##2\endm@tch{%
  \def\tmpa{##1}\def\tmpb{##2}\let\m@tchtmp=\relax%
  \ifx\tmpb\@mpty\def\match{YN}%
  \else\def\match{YY}\fi%
}}%
\def\incm#1{{\psxoffset=1cm\d@my=#1
 \d@mx=\d@my
  \divide\d@mx by \psxoffset
  \xdef\dimincm{\number\d@mx.}
  \advance\d@my by -\number\d@mx cm
  \multiply\d@my by 100
 \d@mx=\d@my
  \divide\d@mx by \psxoffset
  \edef\dimincm{\dimincm\number\d@mx}
  \advance\d@my by -\number\d@mx cm
  \multiply\d@my by 100
 \d@mx=\d@my
  \divide\d@mx by \psxoffset
  \xdef\dimincm{\dimincm\number\d@mx}
}}%
\newif\ifNotB@undingBox
\newhelp\PShelp{Proceed: you'll have a 5cm square blank box instead of
your graphics.}%
\def\s@tsize#1 #2 #3 #4\@ndsize{
  \def\psllx{#1}\def\pslly{#2}%
  \def\psurx{#3}\def\psury{#4}
  \ifx\psurx\@mpty\NotB@undingBoxtrue
  \else
    \drawinght=#4bp\advance\drawinght by-#2bp
    \drawingwd=#3bp\advance\drawingwd by-#1bp
  \fi
  }%
\def\sc@nBBline#1:#2\@ndBBline{\edef\p@rameter{#1}\edef\v@lue{#2}}%
\def\g@bblefirstblank#1#2:{\ifx#1 \else#1\fi#2}%
{\catcode`\%=12
\xdef\B@undingBox{
\def\ReadPSize#1{
 \readfilename#1\relax
 \let\PSfilename=\lastreadfilename
 \openin\pst@mpin=#1\relax
 \ifeof\pst@mpin \errhelp=\PShelp
   \errmessage{I haven't found your postscript file (\PSfilename)}%
   \psloc@lerr{was not found}%
   \s@tsize 0 0 142 142\@ndsize
   \closein\pst@mpin
 \else
   \if\matchexpin{\GlobalInputList}{, \lastreadfilename}%
   \else\xdef\GlobalInputList{\GlobalInputList, \lastreadfilename}%
     \immediate\write\psbj@inaux{\lastreadfilename,}%
   \fi%
   \loop
     \executeinspecs{\catcode`\ =10\global\read\pst@mpin to\n@xtline}%
     \ifeof\pst@mpin
       \errhelp=\PShelp
       \errmessage{(\PSfilename) is not an Encapsulated PostScript File:
           I could not find any \B@undingBox: line.}%
       \edef\v@lue{0 0 142 142:}%
       \psloc@lerr{is not an EPSFile}%
       \NotB@undingBoxfalse
     \else
       \expandafter\sc@nBBline\n@xtline:\@ndBBline
       \ifx\p@rameter\B@undingBox\NotB@undingBoxfalse
         \edef\t@mp{%
           \expandafter\g@bblefirstblank\v@lue\space\space\space}%
         \expandafter\s@tsize\t@mp\@ndsize
       \else\NotB@undingBoxtrue
       \fi
     \fi
   \ifNotB@undingBox\repeat
   \closein\pst@mpin
 \fi
\message{#1}%
}%
\def\psboxto(#1;#2)#3{\vbox{%
   \ReadPSize{#3}%
   \advance\pswdincr by \drawingwd
   \advance\pshtincr by \drawinght
   \divide\pswdincr by 1000
   \divide\pshtincr by 1000
   \d@mx=#1
   \ifdim\d@mx=0pt\xscale=1000
         \else \xscale=\d@mx \divide \xscale by \pswdincr\fi
   \d@my=#2
   \ifdim\d@my=0pt\yscale=1000
         \else \yscale=\d@my \divide \yscale by \pshtincr\fi
   \ifnum\yscale=1000
         \else\ifnum\xscale=1000\xscale=\yscale
                    \else\ifnum\yscale<\xscale\xscale=\yscale\fi
              \fi
   \fi
   \divide\drawingwd by1000 \multiply\drawingwd by\xscale
   \divide\drawinght by1000 \multiply\drawinght by\xscale
   \divide\psxoffset by1000 \multiply\psxoffset by\xscale
   \divide\psyoffset by1000 \multiply\psyoffset by\xscale
   \global\divide\pscm by 1000
   \global\multiply\pscm by\xscale
   \multiply\pswdincr by\xscale \multiply\pshtincr by\xscale
   \ifdim\d@mx=0pt\d@mx=\pswdincr\fi
   \ifdim\d@my=0pt\d@my=\pshtincr\fi
   \message{scaled \the\xscale}%
 \hbox to\d@mx{\hss\vbox to\d@my{\vss
   \global\setbox\drawingBox=\hbox to 0pt{\kern\psxoffset\vbox to 0pt{%
      \kern-\psyoffset
      \PSspeci@l{\PSfilename}{\the\xscale}%
      \vss}\hss\ps@nnotation}%
   \global\wd\drawingBox=\the\pswdincr
   \global\ht\drawingBox=\the\pshtincr
   \global\drawingwd=\pswdincr
   \global\drawinght=\pshtincr
   \baselineskip=0pt
   \copy\drawingBox
 \vss}\hss}%
  \global\psxoffset=0pt
  \global\psyoffset=0pt
  \global\pswdincr=0pt
  \global\pshtincr=0pt 
  \global\pscm=1cm 
}}%
\def\psboxscaled#1#2{\vbox{%
  \ReadPSize{#2}%
  \xscale=#1
  \message{scaled \the\xscale}%
  \divide\pswdincr by 1000 \multiply\pswdincr by \xscale
  \divide\pshtincr by 1000 \multiply\pshtincr by \xscale
  \divide\psxoffset by1000 \multiply\psxoffset by\xscale
  \divide\psyoffset by1000 \multiply\psyoffset by\xscale
  \divide\drawingwd by1000 \multiply\drawingwd by\xscale
  \divide\drawinght by1000 \multiply\drawinght by\xscale
  \global\divide\pscm by 1000
  \global\multiply\pscm by\xscale
  \global\setbox\drawingBox=\hbox to 0pt{\kern\psxoffset\vbox to 0pt{%
     \kern-\psyoffset
     \PSspeci@l{\PSfilename}{\the\xscale}%
     \vss}\hss\ps@nnotation}%
  \advance\pswdincr by \drawingwd
  \advance\pshtincr by \drawinght
  \global\wd\drawingBox=\the\pswdincr
  \global\ht\drawingBox=\the\pshtincr
  \global\drawingwd=\pswdincr
  \global\drawinght=\pshtincr
  \baselineskip=0pt
  \copy\drawingBox
  \global\psxoffset=0pt
  \global\psyoffset=0pt
  \global\pswdincr=0pt
  \global\pshtincr=0pt 
  \global\pscm=1cm
}}%
\def\psbox#1{\psboxscaled{1000}{#1}}%
\newif\ifn@teof\n@teoftrue
\newif\ifc@ntrolline
\newif\ifmatch
\newread\j@insplitin
\newwrite\j@insplitout
\newwrite\psbj@inaux
\immediate\openout\psbj@inaux=psbjoin.aux
\immediate\write\psbj@inaux{\string\joinfiles}%
\immediate\write\psbj@inaux{\jobname,}%
\def\toother#1{\ifcat\relax#1\else\expandafter%
  \toother@ux\meaning#1\endtoother@ux\fi}%
\def\toother@ux#1 #2#3\endtoother@ux{\def\tmp{#3}%
  \ifx\tmp\@mpty\def\tmp{#2}\let\next=\relax%
  \else\def\next{\toother@ux#2#3\endtoother@ux}\fi%
\next}%
\let\readfilenamehook=\relax
\def\re@d{\expandafter\re@daux}
\def\re@daux{\futurelet\nextchar\stopre@dtest}%
\def\re@dnext{\xdef\lastreadfilename{\lastreadfilename\nextchar}%
  \afterassignment\re@d\let\nextchar}%
\def\stopre@d{\egroup\readfilenamehook}%
\def\stopre@dtest{%
  \ifcat\nextchar\relax\let\nextread\stopre@d
  \else
    \ifcat\nextchar\space\def\nextread{%
      \afterassignment\stopre@d\chardef\nextchar=`}%
    \else\let\nextread=\re@dnext
      \toother\nextchar
      \edef\nextchar{\tmp}%
    \fi
  \fi\nextread}%
\def\readfilename{\bgroup%
  \let\\=\backslashother \let\%=\percentother \let\~=\tildeother
  \let\#=\sharpother \xdef\lastreadfilename{}%
  \re@d}%
\xdef\GlobalInputList{\jobname}%
\def\psnewinput{%
  \def\readfilenamehook{
    \if\matchexpin{\GlobalInputList}{, \lastreadfilename}%
    \else\xdef\GlobalInputList{\GlobalInputList, \lastreadfilename}%
      \immediate\write\psbj@inaux{\lastreadfilename,}%
    \fi%
    \let\readfilenamehook=\relax%
    \ps@ldinput\lastreadfilename\relax%
  }\readfilename%
}%
\expandafter\ifx\csname @@input\endcsname\relax    
  \immediate\let\ps@ldinput=\input\def\input{\psnewinput}%
\else
  \immediate\let\ps@ldinput=\@@input
  \def\@@input{\psnewinput}%
\fi%
\def\nowarnopenout{%
 \def\warnopenout##1##2{%
   \readfilename##2\relax
   \message{\lastreadfilename}%
   \immediate\openout##1=\lastreadfilename\relax}}%
\def\warnopenout#1#2{%
 \readfilename#2\relax
 \def\t@mp{TrashMe,psbjoin.aux,psbjoint.tex,}\uncatcode\t@mp
 \if\matchexpin{\t@mp}{\lastreadfilename,}%
 \else
   \immediate\openin\pst@mpin=\lastreadfilename\relax
   \ifeof\pst@mpin
     \else
     \edef\tmp{{If the content of this file is precious to you, this
is your last chance to abort (ie press x or e) and rename it before
retexing (\jobname). If you're sure there's no file
(\lastreadfilename) in the directory of (\jobname), then go on: I'm
simply worried because you have another (\lastreadfilename) in some
directory I'm looking in for inputs...}}%
     \errhelp=\tmp
     \errmessage{I may be about to replace your file named \lastreadfilename}%
   \fi
   \immediate\closein\pst@mpin
 \fi
 \message{\lastreadfilename}%
 \immediate\openout#1=\lastreadfilename\relax}%
{\catcode`\%=12\catcode`\*=14
\gdef\splitfile#1{*
 \readfilename#1\relax
 \immediate\openin\j@insplitin=\lastreadfilename\relax
 \ifeof\j@insplitin
   \message{! I couldn't find and split \lastreadfilename!}*
 \else
   \immediate\openout\j@insplitout=TrashMe
   \message{< Splitting \lastreadfilename\space into}*
   \loop
     \ifeof\j@insplitin
       \immediate\closein\j@insplitin\n@teoffalse
     \else
       \n@teoftrue
       \executeinspecs{\global\read\j@insplitin to\spl@tinline\expandafter
         \ch@ckbeginnewfile\spl@tinline
       \ifc@ntrolline
       \else
         \toks0=\expandafter{\spl@tinline}*
         \immediate\write\j@insplitout{\the\toks0}*
       \fi
     \fi
   \ifn@teof\repeat
   \immediate\closeout\j@insplitout
 \fi\message{>}*
}*
\gdef\ch@ckbeginnewfile#1
 \def\t@mp{#1}*
 \ifx\@mpty\t@mp
   \def\t@mp{#3}*
   \ifx\@mpty\t@mp
     \global\c@ntrollinefalse
   \else
     \immediate\closeout\j@insplitout
     \warnopenout\j@insplitout{#2}*
     \global\c@ntrollinetrue
   \fi
 \else
   \global\c@ntrollinefalse
 \fi}*
\gdef\joinfiles#1\into#2{*
 \message{< Joining following files into}*
 \warnopenout\j@insplitout{#2}*
 \message{:}*
 {*
 \edef\w@##1{\immediate\write\j@insplitout{##1}}*
\w@{
\w@{
\w@{
\w@{
\w@{
\w@{
\w@{
\w@{
\w@{
\w@{
\w@{\string\input\space psbox.tex}*
\w@{\string\splitfile{\string\jobname}}*
\w@{\string\let\string\autojoin=\string\relax}*
}*
 \expandafter\tre@tfilelist#1, \endtre@t
 \immediate\closeout\j@insplitout
 \message{>}*
}*
\gdef\tre@tfilelist#1, #2\endtre@t{*
 \readfilename#1\relax
 \ifx\@mpty\lastreadfilename
 \else
   \immediate\openin\j@insplitin=\lastreadfilename\relax
   \ifeof\j@insplitin
     \errmessage{I couldn't find file \lastreadfilename}*
   \else
     \message{\lastreadfilename}*
     \immediate\write\j@insplitout{
     \executeinspecs{\global\read\j@insplitin to\oldj@ininline}*
     \loop
       \ifeof\j@insplitin\immediate\closein\j@insplitin\n@teoffalse
       \else\n@teoftrue
         \executeinspecs{\global\read\j@insplitin to\j@ininline}*
         \toks0=\expandafter{\oldj@ininline}*
         \let\oldj@ininline=\j@ininline
         \immediate\write\j@insplitout{\the\toks0}*
       \fi
     \ifn@teof
     \repeat
   \immediate\closein\j@insplitin
   \fi
   \tre@tfilelist#2, \endtre@t
 \fi}*
}%
\def\autojoin{%
 \immediate\write\psbj@inaux{\string\into{psbjoint.tex}}%
 \immediate\closeout\psbj@inaux
 \expandafter\joinfiles\GlobalInputList\into{psbjoint.tex}%
}%
\def\centinsert#1{\midinsert\line{\hss#1\hss}\endinsert}%
\def\psannotate#1#2{\vbox{%
  \def\ps@nnotation{#2\global\let\ps@nnotation=\relax}#1}}%
\def\pscaption#1#2{\vbox{%
   \setbox\drawingBox=#1
   \copy\drawingBox
   \vskip\baselineskip
   \vbox{\hsize=\wd\drawingBox\setbox0=\hbox{#2}%
     \ifdim\wd0>\hsize
       \noindent\unhbox0\tolerance=5000
    \else\centerline{\box0}%
    \fi
}}}%
\def\at(#1;#2)#3{\setbox0=\hbox{#3}\ht0=0pt\dp0=0pt
  \rlap{\kern#1\vbox to0pt{\kern-#2\box0\vss}}}%
\newdimen\gridht \newdimen\gridwd
\def\gridfill(#1;#2){%
  \setbox0=\hbox to 1\pscm
  {\vrule height1\pscm width.4pt\leaders\hrule\hfill}%
  \gridht=#1
  \divide\gridht by \ht0
  \multiply\gridht by \ht0
  \gridwd=#2
  \divide\gridwd by \wd0
  \multiply\gridwd by \wd0
  \advance \gridwd by \wd0
  \vbox to \gridht{\leaders\hbox to\gridwd{\leaders\box0\hfill}\vfill}}%
\def\fillinggrid{\at(0cm;0cm){\vbox{%
  \gridfill(\drawinght;\drawingwd)}}}%
\def\textleftof#1:{%
  \setbox1=#1
  \setbox0=\vbox\bgroup
    \advance\hsize by -\wd1 \advance\hsize by -2em}%
\def\textrightof#1:{%
  \setbox0=#1
  \setbox1=\vbox\bgroup
    \advance\hsize by -\wd0 \advance\hsize by -2em}%
\def\endtext{%
  \egroup
  \hbox to \hsize{\valign{\vfil##\vfil\cr%
\box0\cr%
\noalign{\hss}\box1\cr}}}%
\def\frameit#1#2#3{\hbox{\vrule width#1\vbox{%
  \hrule height#1\vskip#2\hbox{\hskip#2\vbox{#3}\hskip#2}%
        \vskip#2\hrule height#1}\vrule width#1}}%
\def\boxit#1{\frameit{0.4pt}{0pt}{#1}}%
\catcode`\@=12 
\psfordvips   

\usepackage{graphics}
\usepackage{multirow}
\usepackage[usenames]{color}
\usepackage{nicefrac}
\usepackage{amsmath}

\newcommand {\mb}[1]{\mbox{\boldmath{${#1}$}}}

\begin{document}

\title{Probing the circulation of ring--shaped Bose--Einstein
condensates }
\author{Noel Murray}
\affiliation{Department of Physics, Georgia Southern University,
Statesboro, GA 30460--8031 USA}
\author{Michael Krygier}
\affiliation{Department of Physics, Georgia Southern University,
Statesboro, GA 30460--8031 USA}
\author{Mark Edwards}
\affiliation{Department of Physics, Georgia Southern University,
Statesboro, GA 30460--8031 USA}
\author{K. C. Wright}
\affiliation{Joint Quantum Institute, National Institute of Standards 
and Technology and the University of Maryland, Gaithersburg, MD 20899, USA}
\author{G. K. Campbell}
\affiliation{Joint Quantum Institute, National Institute of Standards 
and Technology and the University of Maryland, Gaithersburg, MD 20899, USA}
\author{Charles W.\ Clark}
\affiliation{Joint Quantum Institute, National Institute of Standards 
and Technology and the University of Maryland, Gaithersburg, MD 20899, USA}

\date{\today}

\begin{abstract}
This paper reports the results of a theoretical and experimental study of 
how the initial circulation of ring--shaped Bose--Einstein condensates (BECs) 
can be probed by time--of--flight (TOF) images.  We have studied theoretically 
the dynamics of a BEC after release from a toroidal trap potential by solving 
the 3D Gross--Pitaevskii (GP) equation.  The trap and condensate characteristics 
matched those of a recent experiment. The circulation, experimentally 
imparted to the condensate by stirring, was simulated theoretically by 
imprinting a linear azimuthal phase on the initial condensate wave function. 
The theoretical TOF images were in good agreement with the experimental data.  
We find that upon release the dynamics of the ring--shaped condensate proceeds 
in two distinct phases.  First, the condensate expands rapidly inward, filling 
in the initial hole until it reaches a minimum radius that depends on the 
initial circulation.  In the second phase, the density at the inner radius 
increases to a maximum after which the hole radius begins slowly to expand.  
During this second phase a series of concentric rings appears due to the 
interference of ingoing and outgoing matter waves from the inner radius.  The 
results of the GP equation predict that the hole area is a quadratic function of 
the initial circulation when the condensate is released directly from the trap
in which it was stirred and is a linear function of the circulation if the trap
is relaxed before release. These scalings matched the data.  Thus, hole size 
after TOF can be used as a reliable probe of initial condensate circulation.  
This connection between circulation and hole size after TOF will facilitate 
future studies of atomtronic systems that are implemented in ultracold 
quantum gases. 

\end{abstract}

\pacs{03.75.Gg,67.85.Hj,03.67.Dg}

\maketitle

\section{Introduction}
\label{intro}

Recently there has been much research activity devoted to ``atomtronic'' 
systems: confined ultracold atomic--gases that are analogous to electronic 
devices and circuits~\cite{PhysRevLett.103.140405,PhysRevA.75.023615}.  
Atomtronic devices rely on neutral atoms, often Bose--Einstein--condensed, for 
their operation. Characteristics of the atoms in these devices include tunable 
collisional interactions, internal structure, long--range coherence and 
superfluidity.  Thus atom--gas analogs of quite a number of electronic devices 
have been proposed including diodes~\cite{PhysRevA.70.061604,PhysRevLett.100.240407} 
and transistors~\cite{PhysRevA.75.013608}, and a capacitor discharged through 
a resistor~\cite{SciRep.2.352.2013,Esslinger_2012}.  The coherence and superfluid 
properties of these systems make them useful as sensors and other devices that 
can take advantage of superfluidity.  

Of particular interest is the realization of an atomic--gas analog of a 
Superconducting Quantum Interference Device (SQUID).  Traditional SQUIDs are 
used to construct magnetic--field detectors, voltmeters, gradiometers, and a 
host of other metrologic devices~\cite{SQUID_Handbook}.  SQUID circuits have 
been realized with either tunnel or weak--link junctions%
~\cite{RevModPhys.51.101,RevModPhys.74.741}.  
In gaseous Bose--Einstein condensates (BECs), Josephson--like junctions have 
been demonstrated in double--well potentials%
~\cite{PhysRevLett.95.010402,Nature.449.579} 
and a closed--loop atom ``circuit'' was implemented in a ring--shaped confining 
potential~\cite{PhysRevLett.106.130401}.

\begin{figure}[t]
\begin{center}
\mbox{\psboxto(3.0in;0.0in){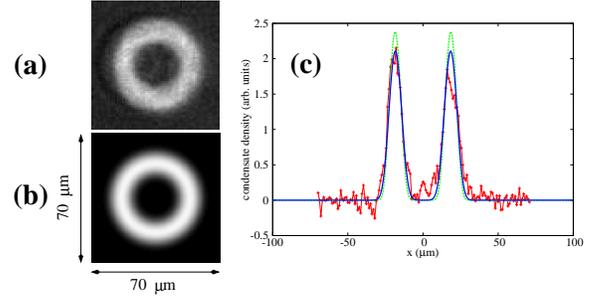}}
\caption{(color online) (a) In--situ 2D density image of the initial condensate 
used to determine the parameters of the confining potential.   (b) The 2D density 
image of the TIGPE theoretical result, and (c) cut--through density profiles 
comparing the experimental data (red curve with points explicitly displayed,) 
the fitted solution of the TIGPE (green, dotted curve,) and the fitted solution 
(blue, solid curve) blurred by convolving it with a Gaussian having a 4--$\mu$m 
1/e$^{2}$ radius.}
\label{figure1}
\end{center}
\end{figure} 

Recently, an atom circuit analogous to an rf SQUID~\cite{SQUID_Handbook} was 
implemented in a ring BEC by creating a rotating weak link (a region of reduced 
superfluid density) with a blue--detuned laser beam~\cite{PhysRevLett.110.025302}.  
In that experiment, the rotating weak link was used to drive phase slips which 
changed the circulation around the ring.  In order to measure the circulation, 
the condensate was released from the trap allowing it to expand in time--of--flight 
(TOF).  The image of the resulting condensate exhibited a smooth density profile 
for zero circulation and a hole whose size depended on the winding number $m$ 
(the number of times the phase winds through $2\pi$ in a closed loop around the 
ring) for non--zero circulation.  The winding numbers of released condensates 
were inferred by measuring the distribution of hole areas in the TOF images.

In this paper we present an experimental and theoretical study of the connection 
between the condensate winding number at the time of release with the size of 
the hole in the TOF image.  This connection has important consequences for the 
field of atomtronics since it provides a solid theoretical foundation for the 
formulation of models that can underly the analogies between electric circuits 
and ultracold atomic--gas systems as seen, for example, in Ref.%
~\cite{PhysRevLett.110.025302}.

The plan of the paper is as follows.  In Section~\ref{experiment} we describe 
the conditions of the experiment where ring--shaped condensates were created, 
stirred, released, and imaged.  Typical results from these experiments are also 
displayed. Section~\ref{model} describes the modeling of the experiment based 
on the Gross--Pitaevskii (GP) equation and Section~\ref{exp_thy_compare} shows 
a comparison of theory and experiment.  We found that theory based on the GP 
equation matched the data well.  Having established that the GP equation can 
provide a good description of condensate behavior under these conditions, we 
present the GP--equation picture of the dynamics of a released ring BEC as a 
function of its initial winding number. Finally, section~\ref{conclusion} 
provides a summary of the results and places the result in context.

\begin{figure}[t]
\begin{center}
\mbox{\psboxto(3.0in;0.0in){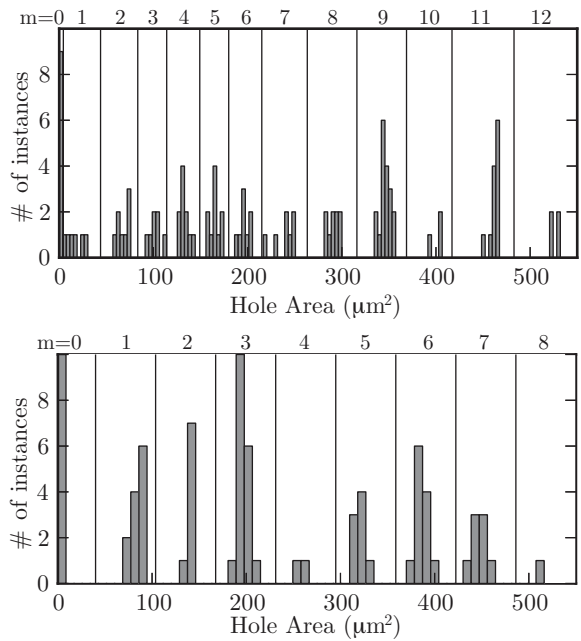}}
\caption{The circulation of stirred condensates was determined experimentally
from the distribution of hole sizes in the expanded cloud.  These distributions 
are shown in the histograms above for the direct--release case (upper) and for 
the ramp--and--release cases (lower). The winding number $m$ assigned for a 
given hole area is labeled for each bin.}
\label{figure2}
\end{center}
\end{figure} 

\section{Description of the experiment}
\label{experiment}

The ring BEC, shown in Fig.~\ref{figure1}a, contains $\approx4\times10^5$ $^{23}$Na 
atoms, spin-polarized in the $3^2S_{1/2}\left|F=1,m_F=-1\right>$ state. The atoms 
are held in an optical dipole trapping potential formed by a combination of up 
to three laser beams. Two of these trapping beams are red-detuned from atomic 
resonance: a horizontal ``sheet'' beam providing the primary confinement in the 
vertical direction with trap frequency $\omega_z/2\pi$ = 600 Hz, and a vertically 
propagating $LG_0^1$ ``ring'' beam, which can be used to create a toroidal potential 
minimum. This red--detuned trap configuration was used previously for experiments 
conducted in the same laboratory~\cite{PhysRevLett.106.130401,PhysRevLett.110.025302}. 
An alternative setup, which uses a vertically propagating blue--detuned ``anti--ring'' 
beam formed by re--imaging the light transmitted through a ring--shaped intensity 
mask onto the BEC is now used to provide a ring--shaped trap during experiments.  
This blue--detuned trap allows the BEC to be confined to a narrower annulus than 
the red trap with a smoother trap minimum. 

As described below, the blue-detuned trap is now used for performing experiments, 
and the red--detuned $LG_0^1$  beam is added only to facilitate loading and releasing 
the blue--detuned ring trap. If the imaging resolution were perfect, the blue-detuned 
trap would be hard-walled, however the 4--$\mu$m resolution of the imaging system 
blurs the edges of the trap. With our narrow annulus, the shadow in the intensity 
pattern can be approximated by a 2D Gaussian ring with a 1/$e^2$ radius of 9(1) 
$\mu$m and mean radius of 18.5(1) $\mu$m.  This smooth, narrow annular trap can 
support stable persistent currents of up to winding number $m=12$ (see below).

\begin{figure*}[t]
\begin{center}
\mbox{\psboxto(7.0in;0.0in){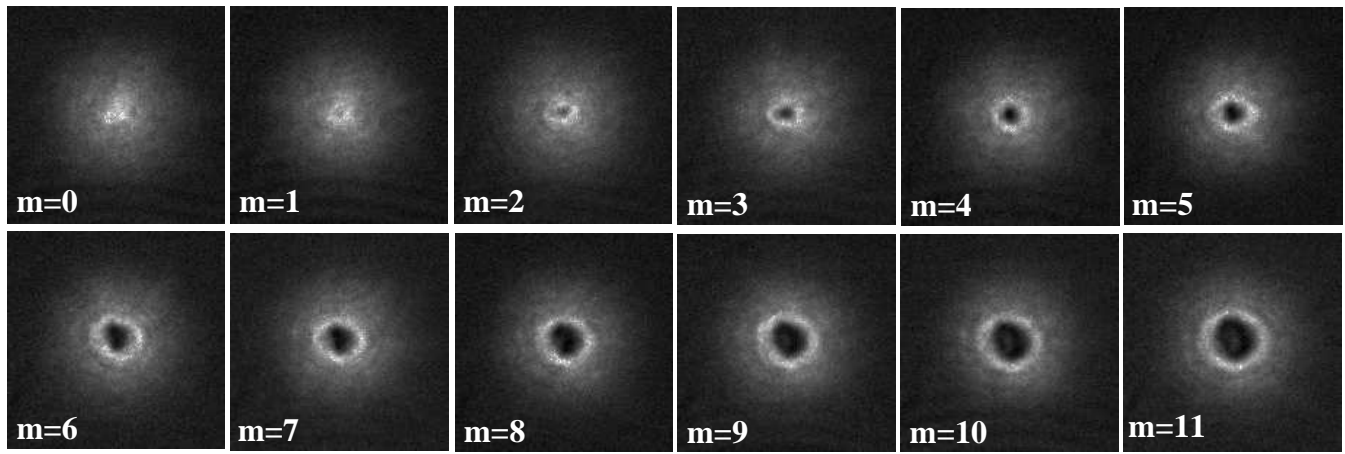}}
\caption{Time--of--flight images of the condensate after direct release from the 
trap and 10 ms TOF.  Each image is labeled with the circulation $m$ that was 
determined by analysis of hole size as shown in Fig.~\ref{figure2}.}
\label{figure3}
\end{center}
\end{figure*} 

For this experiment, the atomic sample was initially trapped and cooled to below 
the BEC transition in the red-detuned ring trap, primarily by ramping down the 
intensity of the sheet beam. After the BEC was formed, the blue--detuned ring was 
ramped up to its full value while keeping the red--detuned ring on after which 
the red-detuned ring was ramped off. The BEC was then driven into a circulating 
state by stirring it with a tightly focused blue--detuned beam propagating in the 
vertical direction. The position and intensity of this beam was controlled by an 
acousto--optic deflector, using the same time-averaged scanning technique reported 
in Ref.~\cite{PhysRevLett.110.025302}, which creates a broad, effectively flat 
potential in the radial direction. 

The height of the (initially stationary) potential barrier was ramped up over 
200 ms to a height above the chemical potential of the BEC. Once at full height, 
the barrier was accelerated in the azimuthal direction around the ring up to a 
final angular velocity $\Omega$, which could be varied between 0 and 15 Hz.  After
this period of acceleration, the velocity was held constant while the barrier 
height was ramped to zero over 200 ms, allowing the condensate to re-connect 
around the ring. 

Immediately following this stirring procedure, the condensate was often in a 
highly excited state with one or more vortex excitations in the annulus. We 
therefore allowed 5.65 seconds of hold time for those excitations to damp out~\cite
{endnote:vortexdamping}, leaving the BEC in a simple, vortex--free persistent 
current state.

\begin{figure*}[t]
\begin{center}
\mbox{\psboxto(6.0in;0.0in){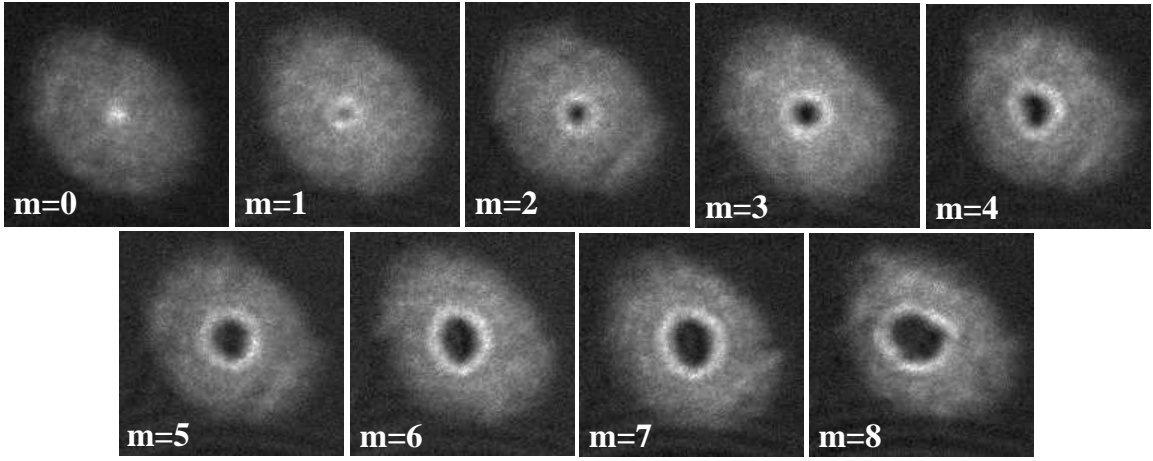}}
\caption{Time--of--flight images of the condensate after ramp--down and release 
from the trap and 10 ms TOF. Each image is labeled with the circulation $m$ that 
was determined by analysis of hole size as shown in Fig.~\ref{figure2}.}
\label{figure4}
\end{center}
\end{figure*} 

Once the BEC was prepared in a circulating state, one of two procedures was 
followed to allow the condensate to expand prior to absorption imaging. In the 
first case, the trapping beams were turned off suddenly ($<$~1 $\mu$s), and the 
condensate was allowed to expand for 10 ms TOF. In the second case, the red--detuned 
ring trap was turned back on, and then the radial confinement of the condensate 
was relaxed to allow atoms to fill in the center of the ring prior to the 10 ms 
TOF. 

This more complex release scheme was developed to make the central hole associated 
with the persistent current visible earlier in TOF expansion. In experiments 
using the blue--detuned trap the hand-off back to the red trap is required because 
lowering of the blue-detuned ring confinement alone allows atoms to escape the 
central trap region into the periphery of the sheet trap. To execute this handoff 
in this experiment, we first turned the red--detuned ring trap back on over 200 ms, 
and then ramped the blue--detuned ring off in 300 ms, and finally ramped the 
red--detuned ring down to about 10\% of its typical power in 50 ms.  Imaging of 
the BEC was done using partial--transfer absorption imaging~\cite
{PhysRevLett.106.130401}. 

A representative set of experimental TOF images is shown in Fig.~\ref{figure3} for 
the direct-release procedure, and in Fig.~\ref{figure4} for the radial relaxation 
procedure. The phase winding number of the persistent current for each run of the 
experiment was determined by measuring the hole sizes for all runs for a given 
release procedure (i.e., either direct--release or radial relaxation) and making 
a histogram plot of these to obtain the distribution of hole sizes.  

There were approximately $100$ experimental runs for each release procedure, as 
shown in Fig.~\ref{figure2}.  The histograms displayed in that figure show that the 
hole sizes cluster around discrete values and that the spaces between these 
clusters exhibit clear gaps.  This hole--size behavior enables the assignment of 
a winding number, $m$, to each cluster starting with $m=0$ for the cluster centered 
at the smallest area and increasing sequentially.  The measured hole area was also 
affected by the imaging resolution, particularly for the direct release images shown 
in Fig.~\ref{figure3} and for small values of $m$ where the 4--$\mu$m resolution is 
comparable to the size of the hole. The stirring speed was varied in such a way 
as to ensure that many samples were obtained for each winding number in the 
expected range. 

\section{Simulating the experiment}
\label{model}

To simulate the experiment we divided it into three distinct phases.  These were 
(1) forming the initial condensate, (2) stirring the condensate to give it a 
non--zero circulation, and (3) probing the condensate by releasing it and allowing 
it to expand before imaging. To simulate phase one we determined the initial 
condensate wave function by solving the time--independent, Gross--Pitaevskii 
equation (TIGPE).  The values of the parameters of the model potential used in 
the TIGPE were determined by varying them until the density profile of the 
solution matched the experimental profile of the {\em in--situ} condensate, as 
shown in Fig.~\ref{figure1}. 

In phase two the result of stirring the condensate was simulated by multiplying 
its wave function by $e^{im\phi}$, where $\phi$ is the azimuthal angle in 
cylindrical coordinates.  This adds $m$ units of circulation to the condensate 
(angular momentum $m\hbar$ per particle if cylindrically symmetric) by imprinting 
a linear azimuthal phase around the ring.  Finally, the release of the condensate 
was simulated by evolving the result of phase (2) using the time--dependent 
Gross--Pitaevskii equation (TDGPE).  Here there were two cases: (3a) direct 
release, where the TDGPE is evolved with $V_{\rm trap}$ set to zero, or (3b) ramp 
and release, where the ring--Gaussian potential depth, $V_{G}$, is reduced to 
about 10\% of its initial value and then the potential is turned off.

The TIGPE used to obtain the initial condensate, $\psi_{0}({\bf r})$, has the 
form,
\begin{equation}
-\frac{\hbar^{2}}{2M}\nabla^{2}\psi_{0} +
V_{\rm trap}({\bf r})\psi_{0} + gN|\psi_{0}|^{2}\psi_{0} = 
\mu_{0}\psi_{0},
\label{tigpe}
\end{equation}
where $M$ is the mass of a ($^{23}$Na) condensate atom, $g$ is the strength of 
the atom--atom interaction due to binary scattering, $N$ is the number of 
condensate atoms, $\mu_{0}$ is the chemical potential and $V_{\rm trap}$ is the 
potential created by external laser beams as described in Sec.\ \ref{experiment}.  

\begin{figure*}[htb]
\begin{center}
\mbox{\psboxto(7.0in;0.0in){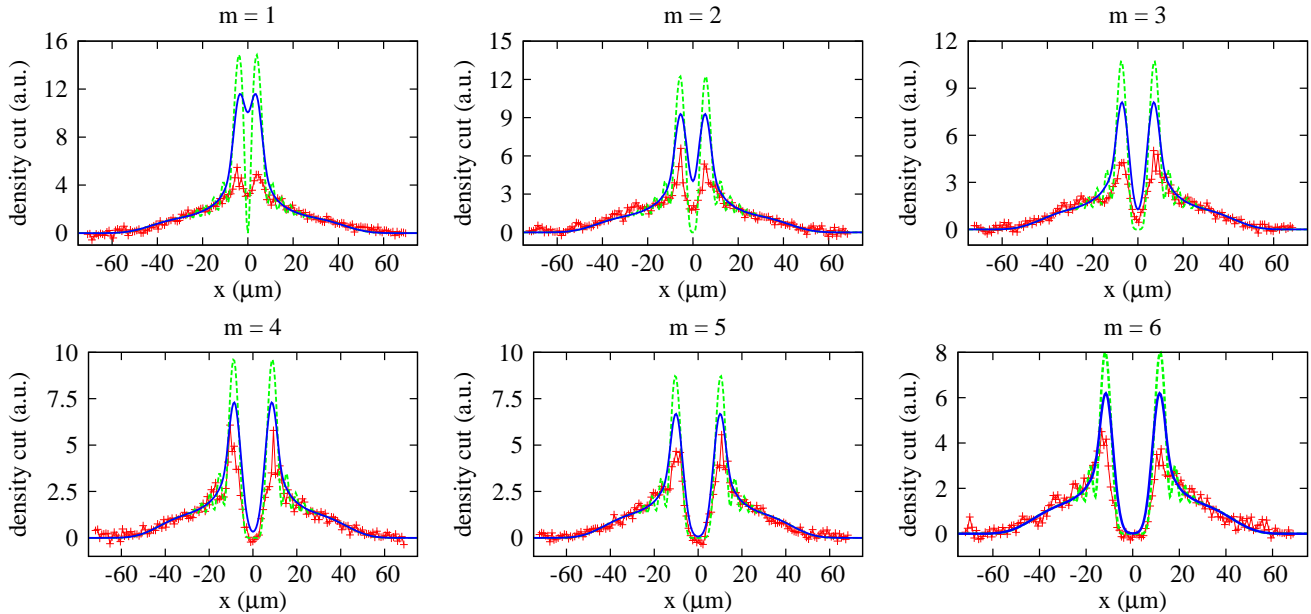}}
\caption{(color online) A comparison of the cut--through of the condensate 
optical density data (shown as individual points connected with a solid red 
line) after direct release and 10 ms TOF with theory for initial circulations 
$m=1,\dots,6$.  Two theoretical curves appear in each graph.  The dotted 
(green) curve is the result of evolving the 3D TDGPE.  The solid (blue) curve 
is a cut--through of the TDGPE prediction convolved with a Gaussian having a 
4--$\mu$m 1/e$^{2}$ radius to account for the finite resolution of the 
experimental instrument.}
\label{figure5}
\end{center}
\end{figure*} 

The potential is modeled as a superposition of a vertical ($z$ direction) 
harmonic potential due to the horizontal light--sheet and a horizontal 
($xy$ plane) ring--Gaussian potential due to the vertical masked--Gaussian 
laser beam.  This potential (in which the energy origin is referenced at the 
potential minimum) has the form
\begin{equation}
V_{\rm trap}({\bf r},t) = 
\tfrac{1}{2}M\omega_{z}^{2}z^{2} + f(t)V_{G}(1-e^{-2(\rho-\rho_{0})^{2}/w^{2}}),
\label{ring_gaussian_potential}
\end{equation}
where $\rho = \sqrt{x^{2}+y^{2}}$ and $f(t)$ is a function of time that 
enables simulation of ramp--and--release--type experiments.  The parameters in 
this potential are $\omega_{z}$, the frequency of the vertical harmonic 
confinement and $V_{G}$, $\rho_{0}$, and $w$, are, respectively, the depth, 
radius, and 1/e$^{2}$ width of the ring--Gaussian potential.

For purposes of numerical work, we introduced a set of scaled units and 
expressed the TIGPE and the TDGPE in terms of scaled variables measured in 
these units.  The scaled units are referenced to a chosen unit of length, 
denoted by $L_{0}$, and scaled spatial coordinates are given by 
$\bar{x}\equiv x/L_{0}$, $\bar{y}\equiv y/L_{0}$, and $\bar{z}\equiv z/L_{0}$.  
Energy and time units are defined in terms of $L_{0}$, $E_{0}\equiv\hbar^{2}/
(2ML_{0}^{2})$ and $T_{0}\equiv\hbar/E_{0}$, enabling the definition of a 
scaled time: $\bar{t}\equiv t/T_{0}$.  Hereafter barred symbols will denote 
quantities expressed in their appropriate scaled units.  

It will also be convenient to express the solution of the TIGPE in terms of scaled 
units as $\psi_{0}({\bf r})\equiv L_{0}^{-3/2}\bar{\psi}_{0}\left({\bf \bar{r}}
\right)$.  In terms of these variables the TIGPE becomes
\begin{equation}
-\bar{\nabla}^{2}\bar{\psi}_{0} + 
\bar{V}_{\rm trap}\bar{\psi}_{0} + 
\bar{g}N\left|\bar{\psi}_{0}\right|^{2}\bar{\psi}_{0}
= \bar{\mu}\bar{\psi}_{0}.
\label{scaled_TIGPE}
\end{equation}
Here $\bar{g}\equiv g/(E_{0}L_{0}^{3})$, $\bar{\nabla}^{2}=(\partial^{2}/
\partial\bar{x}^{2}+\partial^{2}/\partial\bar{y}^{2}+\partial^{2}/\partial\bar
{z}^{2})$ is the Laplacian in terms of scaled variables, and the trap potential 
takes the form:
\begin{eqnarray}
\bar{V}_{\rm trap}(\bar{\bf r},\bar{t}) 
&=&
\bar{\omega}_{z}^{2}\bar{z}^{2} +
f(\bar{t})\bar{V}_{G}(1-e^{-2(\bar{\rho}-\bar{\rho}_{0})^{2}/\bar{w}^{2}})\nonumber\\
&\approx&
\bar{\omega}_{z}^{2}\bar{z}^{2} + 
f(\bar{t})\bar{\omega}_{\rho}^{2}(\bar{\rho}-\bar{\rho}_{0})^{2},
\label{scaled_ring_gaussian_potential}
\end{eqnarray}
where $\bar{\omega}_{z}\equiv\omega_{z}/\omega_{0}$, 
$\bar{\omega}_{\rho}=(2\bar{V}_{G}/\bar{w}^{2})^{1/2}$ and $\omega_{0}$ is the 
scaled frequency unit that is related to the length unit by $L_{0}=(\hbar/M
\omega_{0})^{1/2}$.  The length unit we chose for the results presented here 
was $L_{0}= 10\ \mu$m and, since the condensate was composed of $^{23}$Na atoms, 
the energy unit became $E_{0}=1.444\times10^{-33}\ {\rm J}=0.105\,k_{\rm B}$ nK, 
where $k_{\rm B}$ is Boltzmann's constant.  The time unit was $T_{0}= 72.7$ ms 
and the angular frequency unit was $\omega_{0}=27.5$ rad/s.

To simulate the experiment, the following parameter values were used: $N=400,000$ 
atoms, $\omega_{z}=2\pi\times600$ Hz, $\rho_{0}=18.5\,\mu$m, and $w=9.45\,\mu$m.
These were their experimentally determined values.  This left only the potential 
depth, $V_{G}$, undetermined.  The value of $V_{G}$ was obtained by matching the 
measured density profile of the {\em in--situ} condensate with that predicted by 
the Gaussian--blurred TIGPE solution.  A Gaussian blur was applied to the TIGPE 
solution by convolving it with a Gaussian having a 4--$\mu$m 1/e$^{2}$ width.  
This measured 2D profile is displayed in Fig.~\ref{figure1}(a).  After extracting 
the measured {\em in--situ} density along a cut through the center of the 
initial--state BEC (see Fig.~\ref{figure1}(c)), we solved the TIGPE for various 
values of $V_{G}$ and applied a Gaussian blur to each.  The value of $V_{G}$ 
selected was the one for which the associated Gaussian--blurred TIGPE solution 
best matched the measured cut--through {\em in--situ} density profile of the 
initial condensate.  The fit produced a value of $V_{G}/k_{\rm B}=31.5\,{\rm nK}$ 
(where $k_{\rm B}$ is Boltzmann's constant) for the potential depth.  The 
resulting theoretical 2D density profile is shown in Fig.\ \ref{figure1}(b), and a 
comparison of experimental, TIGPE, and Gaussian--blurred TIGPE cut--through 
densities for the final value of $V_{G}$ is shown in Fig.\ \ref{figure1}(c).  The 
values of $V_{G}$ and $w$ correspond to a radial harmonic frequency of 
$\omega_{\rho}\approx150$ Hz.

The rest of the experiment was simulated by solving the TDGPE, whose form in 
scaled units is
\begin{equation}
i\frac{\partial\bar{\psi}}{\partial\bar{t}} = 
-\bar{\nabla}^{2}\bar{\psi} + 
\bar{U}(\bar{\bf r},\bar{t})\bar{\psi} + 
\bar{g}N\left|\bar{\psi}\right|^{2}\bar{\psi}.
\label{scaled_TDGPE}
\end{equation}
The form of $\bar{U}(\bar{\bf r},\bar{t})$ differed depending on whether 
a direct--release or ramp--and--release experiment was simulated.  This will 
be discussed more fully below.  The TDGPE was solved using the split--step 
Crank--Nicolson algorithm~\cite{Muruganandam20091888} on a 3D $xyz$ spatial 
grid of 400$\times$400$\times$200 points which gridded a box of dimensions 
20$\times$20$\times$10 scaled length units.  This translated into a box of 
200$\times$200$\times$100 $\mu$m, which proved to be large enough to completely 
eliminate wall-bounce effects. The time--of--flight (TOF) of condensate 
expansion was 10 ms in both the direct--release and ramp--and--release cases.  
The TIGPE solution used to fit the potential parameters was obtained by solving 
the TDGPE in imaginary time using the same algorithm and grid characteristics.

\begin{figure*}[htb]
\begin{center}
\mbox{\psboxto(7.0in;0.0in){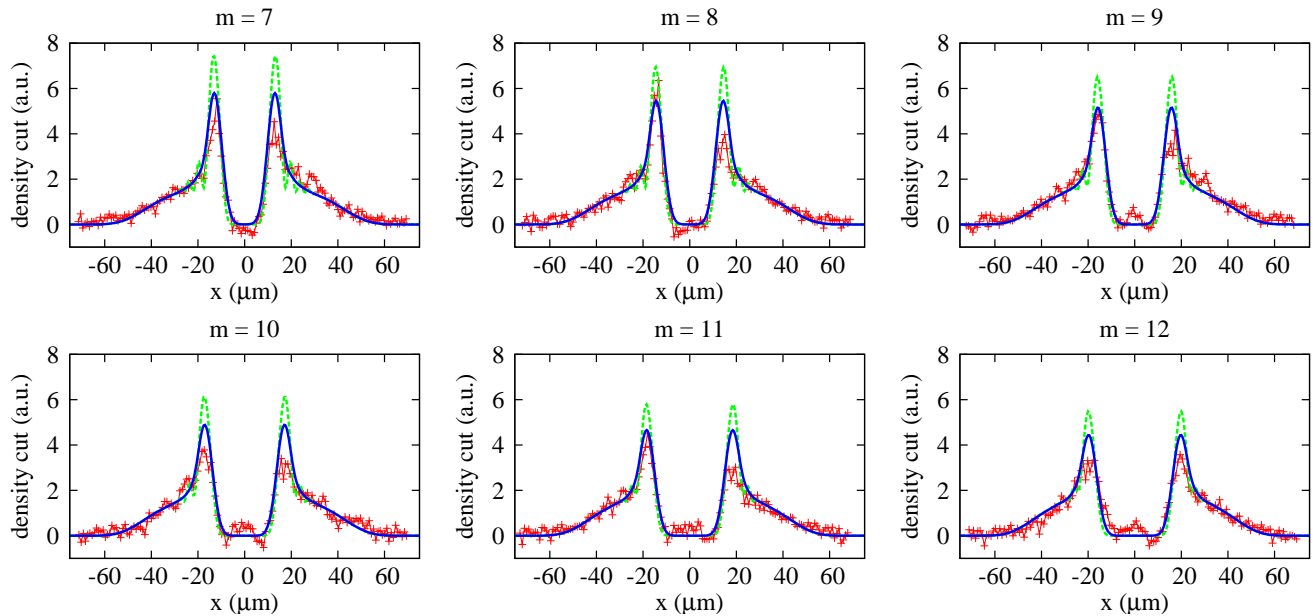}}
\caption{(color online) A comparison of the cut--through of the condensate 
optical density data after direct release and 10 ms TOF with theory for initial 
circulations $m=7,\dots,12$.}
\label{figure6}
\end{center}
\end{figure*} 

\section{Experiment/Theory Comparison}
\label{exp_thy_compare}

\subsection{Direct Release}
\label{direct}

The comparison of theory and experiment for the direct--release case is shown 
in Figs.~\ref{figure5} and~\ref{figure6}.  Figure \ref{figure5} depicts the comparison 
of condensates which were directly released after $1\le m\le 6$ units of angular 
momentum have been applied.  As noted above, for the theory curves angular 
momentum is added to the initial condensate via phase imprint.  The initial 
angular momentum corresponding to the experimental data was determined by 
analysis of the sizes of the holes in the final image.  

Each of the six graphs displays cut--throughs of the optical density of the 
condensate cloud after TOF.  In addition to the data (red curves with individual 
points visible), two theory curves are plotted. The dashed (green) line is a 
cut--through of the density predicted by the TDGPE.  The solid (blue) line is a 
cut--through of the TDGPE prediction convolved with a Gaussian with a 4--$\mu$m 
$1/e^{2}$ radius.  Applying this average to the theoretical results accounts for 
the finite resolution of the imaging optics in the NIST experiment.  Figure 
\ref{figure6} displays the identical comparison for the cases $7\le m\le 12$.

To create the comparisons shown in Figs.\,\ref{figure5} and \ref{figure6} both the 
experimental data and the theoretical results were subjected to some post 
processing.  Extracting the experimental cut--through from the raw images 
involved comparing cut--through data along different lines through each image.  
The most symmetric and highest amplitude density data was selected as the best 
representative data for each image. This amounted to an estimate of the location 
within each image of the center of the trap.  When creating the experiment/theory 
comparison plots, we shifted the experimental data left--right so that the center 
of the trap in the data coincided with the theoretical origin of coordinates.  

We also processed the theory results by multiplying them by an overall normalization 
constant.  This was necessary because the raw experimental data was expressed in 
arbitrary units and the normalization of the theory amounted to a units conversion.  
We determined a normalization constant, $N_{\rm dr}$, for the direct--release 
case by fitting the tail of the $m=1$ density profile.  This normalization constant 
was then applied to the theoretically determined density profiles for all other 
$m$ values in the direct--release case without further adjustment.  The same 
procedure of fitting the tail of the $m=1$ density profile was followed in the 
ramp--and--release case to obtain a normalization constant, $N_{\rm rr}$, and 
this value was applied to the density profiles of all other $m$ values.  The 
values of $N_{\rm dr}$ and $N_{\rm rr}$ were not the same.

As seen in Figs.\,\ref{figure5} and \ref{figure6}, direct release and subsequent 
expansion of a stirred ring BEC results in an annulus--shaped optical density 
image when viewed from above.  One can see that the agreement between theory and 
experiment is very good for directly released condensates.  This is especially 
the case when the instrument resolution is accounted for.  For all $m$ values, 
the size of the hole increases with initial angular momentum and the 
experimental/theory agreement for the hole size is especially good.  We also 
find good agreement out in the wings of the distribution.  Peak--height agreement 
is not as good, however, this may be due to variations in potential--well depth 
around the ring.

\begin{figure*}[htb]
\begin{center}
\mbox{\psboxto(7.0in;0.0in){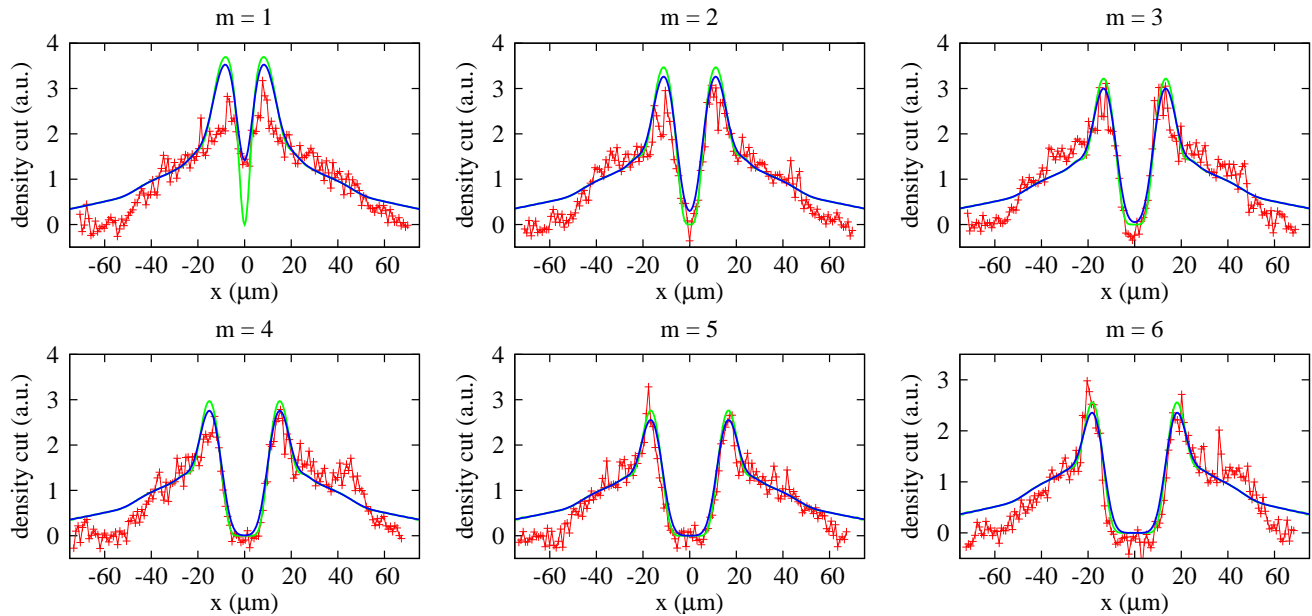}}
\caption{(color online) A comparison of the cut--through of the condensate optical 
density data after ramp--down and release and 10 ms TOF with theory for initial 
circulations $m=1,\dots,6$.}
\label{figure7}
\end{center}
\end{figure*} 

\subsection{Ramp and Release}
\label{ramp}

To model the ramp--down, the potential depth of the initial trap was ramped down 
to 10\% of its initial value over 50 ms. The condensate was then released and 
allowed to expand for 10 ms TOF.  The experiment/theory comparison for this case 
is shown in Fig.\ \ref{figure7}.  This figure contains six comparison graphs in the 
same format as in Figs.\ \ref{figure5} and \ref{figure6} where the initial angular 
momentum ranges over $1\le m\le 6$. The experimental data and theoretical results 
were processed in the same way as for the direct release cases.

For the ramp--down case we find that there is less good agreement in the tails of 
the distribution than in the direct--release case but that peak sizes and especially 
hole sizes match quite well.  It is also clear that the sizes of the holes are 
larger for the same $m$ value than in the direct--release case. Thus ramping down 
the strength of the radial confinement provides a better experimental signature 
of the initial circulation than direct release.

The agreement between the experimental data and the predictions of the TDGPE 
indicates that mean--field theory is adequate to understand how hole sizes in 
the expanded cloud can be used as signatures of condensate circulation before 
release.  Thus it follows that we should be able to rely on the TDGPE to also be 
a reliable predictor of the dynamics of the condensate cloud after release.  
This conclusion may be contrasted with recent work~\cite{PhysRevA.80.021601,%
arxiv.org_abs_1207.0501,0953-4075-46-9-095302},\cite{endnote:hysteresis_paper} 
which indicates that some of the dynamics of the {\em in--situ} cloud (e.g., 
phase--slippage due to vortex dynamics) may not be as well predicted by zero--%
temperature mean--field theory.

\section{Behavior of the released ring BEC}
\label{dynamics}

\subsection{Expansion dynamics}
\label{expansion}

\begin{figure*}[htb]
\begin{center}
\mbox{\psboxto(7.0in;0.0in){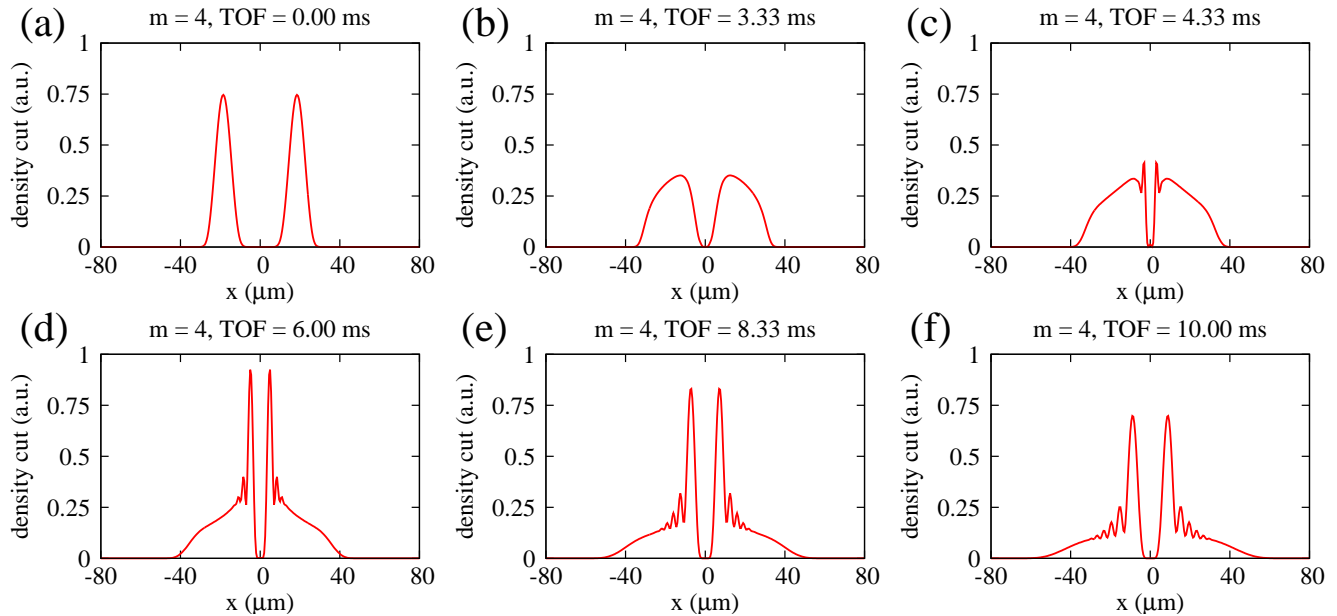}}
\caption{(color online) A sequence of cut--throughs of the optical density as a 
function of time after direct release depicting the dynamics of the expansion.
The initial condensate was phase--imprinted with four units of angular momentum.}
\label{figure8}
\end{center}
\end{figure*}

The dynamical behavior of a released ring condensate is illustrated in 
Figs.\,\ref{figure8}(a)--(f). These figures consist of a sequence of cut--throughs 
of the optical density as a function of time after release. Four units of angular 
momentum ($m=4$) has been added to the initial BEC.  At the moment of release the 
condensate has the shape of the initial ring (Fig.\ \ref{figure8}(a)).  After release 
the condensate expands rapidly inward and more slowly outward.  The inward 
expansion proceeds until the hole reaches a minimum radius (Fig.\ \ref{figure8}(b)).  
When the hole reaches its minimum size (which depends upon the initial angular 
momentum of the condensate) the density around the edge of the hole begins to 
increase rapidly as seen in Fig.\,\ref{figure8}(c).  As the density around the edge 
grows toward its maximum height, a series of density rings begins to form 
(Fig.\,\ref{figure8}(d)).  As the peak density begins slowly to decrease, the size 
of the central hole begins increasing and more rings form as seen in 
Figs.\,\ref{figure8}(e) and \ref{figure8}(f).  In the limit of large times (not shown), 
the peak flattens out and the condensate shape takes the form of a large central 
hole surrounded by a density plateau which is, in turn, surrounded by a series 
of density rings.

The appearance of rings during the rapid increase of the density around the edge 
of the hole seems to be an interference effect.  Condensate atoms flow in towards 
the edge of the hole and then flow out as evidenced by the rapid increase and 
subsequent slower decrease of the density maximum.  At a fixed distance (but 
larger than the hole radius) from the center of the condensate, condensate atoms 
can arrive there by two distinct pathways.  Either they are flowing in towards 
the edge of the hole or they are flowing back out having already visited the hole 
edge.  These two possibilities create two pathways to the same location and thus 
exhibit quantum interference.  

This dynamical behavior also highlights an essential difference between simply 
and multiply connected condensates.  In simply connected condensates the velocity 
distribution can generally be probed by release and subsequent imaging of the 
density profile.  This works as long as interactions between condensate atoms 
during the expansion can be neglected.  It is clear that this is not the case 
for ring condensates.

\subsection{Hole size scaling with winding number}
\label{hole_sizes}

\begin{figure}[htb]
\begin{center}
\mbox{\psboxto(3.0in;0.0in){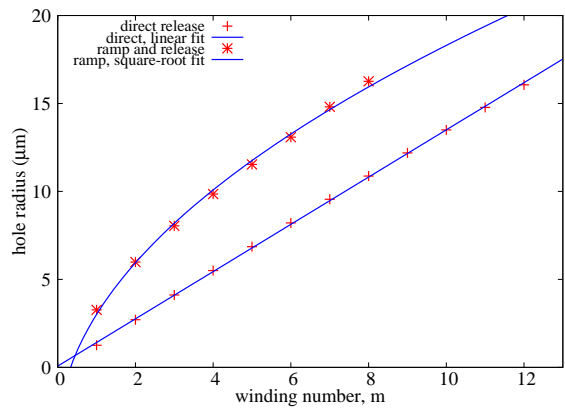}}
\caption{(color online) Hole size scaling as a function of initial winding 
number as predicted by the TDGPE.  The hole radius was defined by the point 
at which the optical density reached 40\% of its peak value closest to 
the trap axis. The lower curve displays hole radii for the direct--release 
simulations (plus symbols).  The solid line appearing in the lower curve is 
a fit to the direct--release radii to the function $am+b$.  The upper curve 
shows corresponding results for the ramp--and--release case.  The asterisks
symbols depict the radiii while the solid curve is a fit of the ramp--and--release
hole radii to the function $\alpha\sqrt{m}+\beta$.}
\label{figure9}
\end{center}
\end{figure}

It is interesting to compare the theoretical predictions for the scalings of 
the hole sizes with initial winding numbers for the direct--release case with
the ramp--and--release case.  Figure\,\ref{figure9} displays the scalings for 
the two cases.  The lower curve in that figure shows the radii, $R_{\rm direct}
(m)$, of the holes versus winding number for the direct--release case for 
$m=0,\dots,12$.  The hole radius was defined to be 40\% of the peak density 
and this percentage was chosen so that the extrapolation of the radii back to
$m=0$ went through zero.  This choice of percentage did not affect the scaling
of the radii with $m$.  The solid line that also appears in the lower curve 
of Fig.\,\ref{figure9} is a linear fit ($R_{\rm direct}(m)=am+b$) to the hole radii.  
The values of the fit parameters were $a=1.34268\,\mu$m and $b=0.0732567\,\mu$m.
Thus hole radii for the direct--release case appear to scale linearly with the
initial circulation of the ring BEC over the entire range of $m$.

The ramp--and--release case shows a different scaling and the radii versus $m$, 
$R_{\rm ramp}(m)$, are shown in the upper curve of Fig.\,\ref{figure9}.  Here the
same definition of hole radius was used.  However, the radii do not appear to
be linear.  This conclusion is supported by the accompanying solid line which
is a fit to a linear function of the {\rm square--root} of $m$: $R_{\rm ramp}(m)
=\alpha\sqrt{m}+\beta$.  The result of the fit yielded $\alpha=7.06904\,\mu$m
and $\beta=-4.05276\,\mu$m.

One effect of these different scalings is that the hole {\em areas} for the
direct--release case will scale quadratically with $m$ while the hole areas
for the ramp--and--release case will scale linearly with $m$.  Evidence of 
this effect can be seen in Fig.\,\ref{figure2} where histograms of the experimental
hole areas are shown.  In Fig.\,\ref{figure2}(a), which shows the direct--release
hole areas, the intervals between hole--area--clusters can be seen to grow
larger as $m$ increases.  In the ramp--and--release case, shown in Fig.\,%
\ref{figure2}(b), these intervals seem to stay the same as $m$ increases.  This 
supports the conclusion that direct--release hole areas scale as $m^{2}$ while 
ramp--and--release areas scale as $m$.

\section{Conclusion}
\label{conclusion}

In this paper we have shown by comparing theory and experiment how the circulation 
of a stirred ring Bose--Einstein condensate can be probed by measuring the size 
of the hole after release and expansion.  We found that the experimental data and 
the predictions of the time--dependent Gross--Pitaevskii equation for the sizes of 
holes produced by stirred and released condensates agree well over a large range 
of winding numbers.  This was the case for condensates released directly after 
stirring and also after ramp down.

We also used the agreement with the TDGPE to understand the dynamics of the 
condensate after release.  We found that the hole of the ring is initially filled 
in to a minimum radius that depends on the winding number at release time.  
Density around the edge of the minimum--radius hole increases to a maxmimum and 
then begins to decrease.  Just after the time of maximum density interference 
rings begin to appear.  Since atoms flow in and then flow out, the rings seem 
to be due to interference between atoms that are flowing in and those flowing 
out.

Ring condensates have exhibited phenomena such as persistent currents and 
deterministic phase slips which came as the result of stirring~\cite
{PhysRevLett.110.025302,PhysRevLett.106.130401}.  In the latter case, the system 
was modeled as an ``atom circuit'' in which the phase changes that occur when 
traversing a closed path around the ring sum to a definite and known quantized 
value.  Key to evaluating these models is the ability to determine the winding 
experimentally and correlating observed condensate behavior with that predicted
by the model.  This paper provides a firm foundation for making these connections.  
Such models will be essential in understanding the behavior of future ``atomtronic'' 
systems based on Bose--Einstein condensates confined in ring potentials.

\begin{acknowledgments}
This work was supported by the Office of Naval Research, the National Science 
Foundation under Physics Frontiers Center grant PHY-0822671 and also by grant 
PHY--1068761, Army Research Office Atomtronics MURI, and the National Institute 
of Standards and Technology.
\end{acknowledgments}

\bibliography{release_dynamics_paper_revised}{}

\begin{thebibliography}{20}
\bibitem[{end({\natexlab{a}})}]{endnote:vortexdamping}
\bibinfo{note}{The decay time for vortex excitations in the annulus varies with
  experimental conditions, but is typically $\approx$3 seconds.}
\end{thebibliography}

\end{document}